\documentclass{article}
\usepackage{spconf,amsmath,graphicx}
\usepackage{xcolor}
\usepackage{multirow}
\usepackage{comment}
\usepackage{url}
\usepackage{hyperref}
\usepackage{enumitem}
\usepackage{enumitem}
\setlist{nosep, leftmargin=14pt}
\usepackage[tableposition=above]{caption}

\usepackage{booktabs}
\usepackage{arydshln}

\usepackage{csquotes}

\title{Polyp and Surgical Instrument Segmentation\\ with Double Encoder-Decoder Networks}

\name{Adrian Galdran$^{1}$}
 \address{$^{1}$ BCN MedTech, Departament de Tecnologies de la Informació i les Comunicacions, \\Pompeu Fabra University, Barcelona, Spain}

\begin{document}
\maketitle
\begin{abstract}
This paper describes a solution for the MedAI competition, in which participants were required to segment both polyps and surgical instruments from endoscopic images. Our approach relies on a double encoder-decoder neural network which we have previously applied for polyp segmentation, but with a series of enhancements: a more powerful encoder architecture, an improved optimization procedure, and the post-processing of segmentations based on tempered model ensembling. Experimental results show that our method produces segmentations that show a good agreement with manual delineations provided by medical experts..
\end{abstract}

\section{Introduction}
Examination of the human gastrointestinal tract is key for the early detection and treatment of different diseases like Colorectal Cancer \cite{haggar_colorectal_2009}. This is usually achieved by a colonoscopic screening, a procedure in which a flexible tube equipped with a camera is introduced through the rectum to look for, and potentially remove lesions throughout the colon. In this context, automated endoscopic image analysis and decision support systems based on artificial intelligence have shown great potential for improving examination effectiveness and facilitating medical intervention \cite{lui_new_2020,jha2021exploring}.

This paper describes our solution to the MedAI medical image segmentation competition \cite{MediAI2021}, a public contest in which participants were asked to solve two independent tasks, namely polyps and surgical instrument extraction from endoscopic images. For this, we employed a deep neural network architecture based upon previous work of us that won the Endotect 2021 challenge \cite{galdran2021double,hicks_endotect_2021}. 
Here we adopt several improvements over our previous polyp segmentation solution, and apply it also for surgical instrument segmentation, as described in the next section. We also experiment with temperature scaling and analyze its impact on resulting performance metrics.

\section{Materials and methods}
Semantic segmentation is typically approached with encoder-decoder networks \cite{ronneberger_u-net_2015} that produce pixel-wise probabilities. 
The encoder acts as a feature extractor, downsampling spatial resolutions and increasing the number of channels by learning convolutional filters. 
The decoder then upsamples this compressed representation back to the original input size. 
Double encoder-encoders are a direct extension in which two encoder-decoder networks are sequentially combined \cite{galdran_little_2020}.

\begin{figure*}[htp]
  \centering
  \includegraphics[width=2\columnwidth]{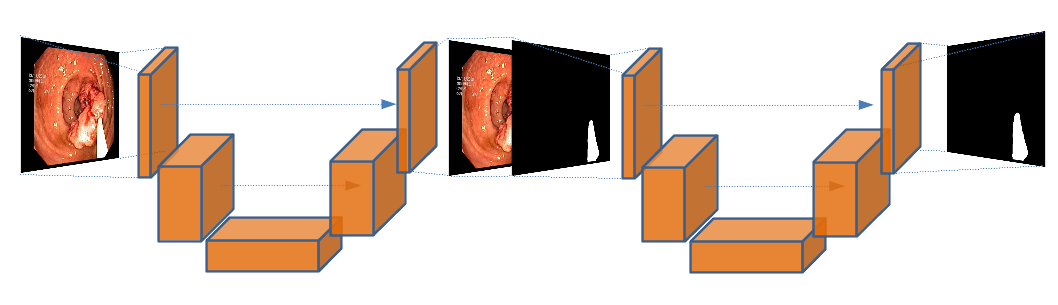}
  \caption{Graphical description of the adopted double encoder-decoder architecture for polyp and instrument segmentation.\label{fig_general}}
\end{figure*}

With $x$ an input RGB image, $E^{(1)}$ the first network, and $E^{(2)}$ the second network, in a double encoder-decoder, the output $E^{(1)}(x)$ of the first network is fed to the second network together with $x$, behaving like a relevance map allowing $E^{(2)}$ to focus on the most interesting parts of the image:
\begin{equation}\label{wnet_def}
E(x) = E^{(2)}(x, E^{(1)}(x)),
\end{equation}
where $x$ and $E^{(1)}(x)$ are stacked so the input to $E^{(2)}$ has four channels, as illustrated in Fig. \ref{fig_general}.
Here we employ the same structure in both $E^{(1)}$ and $E^{(2)}$:a Feature Pyramid Network architecture as the decoder \cite{lin_feature_2017}, and Resnext101 as the pretrained decoder \cite{kolesnikov2020big}. 
In addition, we utilize SAM wrapping ADAM as the optimization procedure \cite{foret_sharpness-aware_2021}, and rotate the train/validation data in a 4-fold manner, resulting in four models $E_1,E_2,E_3,E_4$. We then apply temperature sharpening on the resulting ensemble \cite{NEURIPS2019_1cd138d0}:
\begin{equation}
p = \frac{1}{4} \displaystyle \sum_{i=1}^4 (E_i(x)^t
\end{equation}
with $t$ a free parameter, to obtain our final predictions.

\begin{table*}[!h]  %
	\renewcommand{\arraystretch}{1.4}	
	\centering
\setlength\tabcolsep{5pt}	
\begin{tabular}{l ccccccccc}
 \textbf{Temperature} $\boldsymbol{\rightarrow}$ & \multicolumn{3}{c}{\textbf{T=0.5} } & \multicolumn{3}{c}{\textbf{T=1}} & \multicolumn{3}{c}{\textbf{T=2}} \\
 \cmidrule(lr){1-1} \cmidrule(lr){2-4} \cmidrule(lr){5-7} \cmidrule(lr){8-10} 
                        &  DICE &  PRECISION &  RECALL   &  DICE &  PRECISION  &    RECALL   &  DICE  &   PRECISION  &  RECALL     \\
\midrule
\textbf{Polyps}         & 88.59 & 87.76 &  \textbf{94.56}  &  \textbf{89.65} & 92.42 &  90.09    &  88.69 &  \textbf{93.37} & 87.56 \\\hdashline[2pt/5pt]   %

\textbf{Instruments}    & 94.94 & 92.08 &  \textbf{98.84}  &  96.18 & 95.06 &  97.88    &  \textbf{96.35} &  \textbf{96.32} & 96.92 \\                      %
\bottomrule\\[-0.50cm]
\end{tabular}
\caption{Performance analysis, in terms of Dice score, precision, and recall, of different Temperature sharpening values, for the tasks of polyp segmentation (top row) and instrument segmentation (bottom row).}
\label{tab_results1}
\end{table*}

\section{Experimental Results}
The main metric of interest in medical image segmentation tasks is usually the Dice score, given by:
\begin{equation}\label{dice}
D(f_\theta(x),y) = \frac{\displaystyle 2 \cdot |f_\theta(x) \wedge y|}{|f_\theta(x)| + |y|},
\end{equation}
where $f_\theta(x)$ is the model's binary prediction for example $x$ and $y$ is its corresponding manual ground-truth. If we denote True Positives as $\textrm{TP}$, False Negatives as $\textrm{FN}$, etc., we see that in the numerator of eq. (\ref{dice}), $f_\theta(x) \wedge y$ corresponds to $\textrm{TP}$. 
On the other hand, the denominator describes the amount of pixels that are predicted as foreground ($|f_\theta(x)|=\textrm{TP} + \textrm{FP}$) and the amount of pixels that actually belong to the foreground ($|y| = \textrm{TP} + \textrm{FN}$). 
Therefore we can re-write the Dice score as $D=2\textrm{TP}/(2\textrm{TP}+\textrm{FP}+\textrm{FN})$, and realize that True Negatives have no impact on this metric. 
This means that correctly predicting small objects will be rewarded much more than correctly producing good segmentations of large objects. 
We also report in Table \ref{tab_results1} Precision and Recall:
\begin{equation}
Precision = \frac{\displaystyle  \textrm{TP}}{\displaystyle  \textrm{TP}+ \textrm{FP}}, \ \ \ \  
Recall = \frac{\displaystyle  \textrm{TP}}{\displaystyle  \textrm{TP}+ \textrm{FN}}.
\end{equation}
Achieving a high precision can be considered akin to a good Dice score, while recall also encompasses False Negatives and captures also performance on segmenting large objects.

Table \ref{tab_results1} shows the performance of our approach for the two tasks of Polyp and surgical instrument segmentation on the Kvasir-SEG \cite{jha2020kvasir} and Kvasir-Instrument \cite{Jha2020} datasets. Note that performance is computed in a hidden test set by the organizers of the MedAI competition.

\section{Discussion of the Results and Future Work}
We conclude several observations from the above results. First, instrument segmentation is relatively easier than polyp extraction, due to the greater variability of the latter regarding color and appearance. Also, in both cases we achieve a Dice score close to or above $90\%$, a good overall scoring. Last, temperature has a noticeable impact in performance: when a model attains greater Dice score, it also achieves higher Precision but lower Recall. It can be seen that lower temperature values lead to higher recall at the expense of a decrease in Dice and Precision, whereas higher temperatures result in greater Dice/Precision, but lower Recall. Therefore, Temperature sharpening is a reasonable mechanism to deal with the Precision/Recall trade-off in segmentation tasks.

Although the training data contained images in which there was always a polyp or a surgical instrument present, the test set had frames with no object of interest to be segmented, which is an example of out-of-distribution data. Future research may involve the investigation of different approaches to better handling this kind of problem for the task of endoscopic image segmentation.

\subsection{Acknowledgments}
This project has received funding from the European Union's Horizon 2020 research and innovation programme under the Marie Sklodowska-Curie grant agreement No. 892297.
The author is grateful to {\emph datacrunch.io} for donating GPU time for this project.

\bibliographystyle{IEEEbib}
\bibliography{medai_arxiv}

\end{document}